\documentclass[letterpaper]{article}

\usepackage{aaai}
\usepackage{times}
\usepackage{helvet}
\usepackage{courier}
\usepackage{multirow}
\usepackage{balance}  
\usepackage{graphics} 
\usepackage{times}    
\usepackage{url}      
\usepackage{balance}  
\usepackage{graphicx}
\usepackage{array}
\usepackage{makecell}

\newcolumntype{"}{@{\hskip\tabcolsep\vrule width 1pt\hskip\tabcolsep}}
\makeatother

\begin{document}


 \title{Like Partying? Your Face Says It All.\\
 Predicting the Ambiance of Places with Profile Pictures}
\author{Miriam Redi \\ Yahoo Labs\\\textit{redi@yahoo-inc.com} \And Daniele Quercia\\University of Cambridge\\\textit{dquercia@acm.org}  \And Lindsay T. Graham\\University of Texas, Austin\\\textit{lindsaytgraham@gmail.com}  \And Samuel D. Gosling\\University of Texas, Austin\\\textit{samg@mail.utexas.edu} }
\maketitle

\begin{abstract}
To choose restaurants and coffee shops, people are increasingly relying on social-networking sites.
In a popular site such as Foursquare or Yelp, a place comes with descriptions and reviews, and with  profile 
pictures of people who frequent them. Descriptions and reviews have been widely explored in the research area of data mining. By contrast, profile pictures have received little attention. Previous work showed that people are  able to partly guess a place's ambiance, clientele, and activities not only by observing the place itself but also by observing the profile pictures of its visitors. Here we further that work by determining which visual cues people may have relied upon to make their guesses; showing that a state-of-the-art algorithm could make predictions more accurately than humans at times; and demonstrating that the visual cues people relied upon partly differ from those of the algorithm.
\end{abstract}

%
%
%

\begin {figure*}
\centering
\includegraphics[width=0.7\textwidth ]{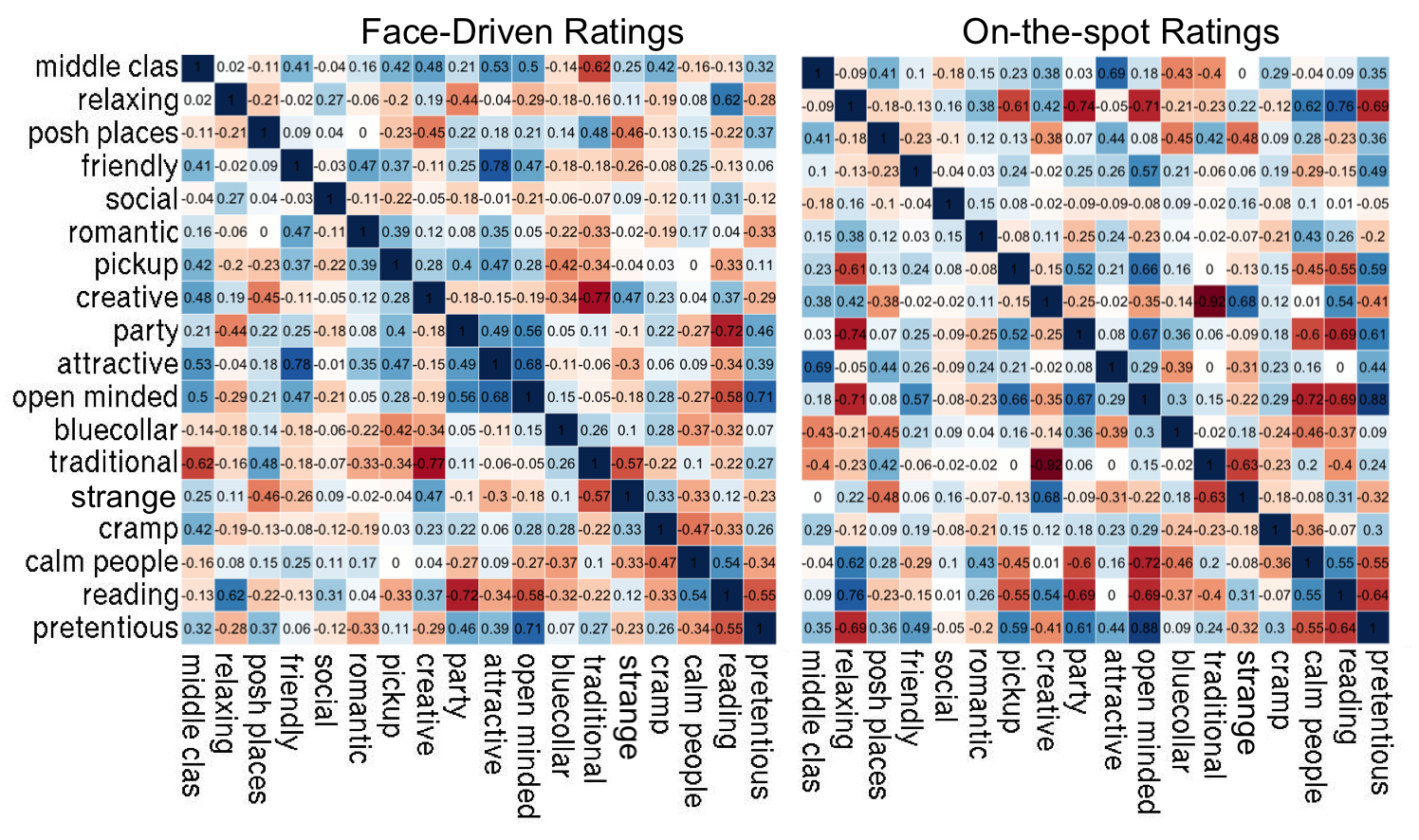}\caption{Correlation matrix of the target ambiance ratings. }
\label{fig:correlation_matrixes}
\end {figure*}

\section{Introduction}


The Internet is going local. Location-based sites like Foursquare are becoming  local search engines, in that, they recommend places based on where users (and their friends) have been in the past. State-of-the-art data mining tools  produce those recommendations by automatically analyzing ratings and reviews~\cite{Vasconcelos:2014:PDF:2660460.2660484}.

As we shall see in Section~\ref{sec:related}, those tools are well-established  and make numbers (ratings) and pieces of text (reviews) relatively easy to mine. By contrast, mining pictures has been proven somewhat harder.  Most of the computer vision research  has been active in making  algorithms more accurate. One of its subareas is called computational aesthetics and, interestingly, is concerned with proposing new ways of automatically extracting visual features that are good proxies for abstract concepts such as beauty and creativity~\cite{redi6}. It comes as no surprise that,  being only at their early stages, computation aesthetics algorithms have not been widely used on social-networking sites.


Here we set out to study whether social-networking sites such as Foursquare might benefit from 
analyzing pictures with computation aesthetics techniques. To determine `for what' pictures might be useful, consider the work done by~\cite{graham11ambience}. The two researchers showed that  people are able to  guess place ambiance  (e.g., whether a restaurant is romantic, whether a coffee shop is friendly) by looking at the profile pictures of visitors. They did so by comparing two types of scores: ambiance ratings given by survey respondents who looked only at profile pictures of visitors; and  ambiance ratings given by study participants who actually visited the places. That work showed that people are partly able to determine the ambiance of a place only by looking at the profile pictures of its visitors. It did not show, however, which visual cues the respondents may have relied upon to make their guesses.

Our goal is to determine whether  state-of-the-art vision techniques could automatically infer place ambiance. In so doing, we make six main contributions by:

\begin{itemize}
\item Analyzing the sets of ambiance ratings collected by~\cite{graham11ambience} (Section~\ref{sec:dataset}). We find that, by considering the pairwise correlations between the 72 ambiance dimensions, one can group those dimensions into 18 fairly orthogonal ones. 

\item Implementing a variety of state-of-the-art computer vision tools (Section~\ref{sec:predictors}). These tools extract the visual cues that the literature of computational aesthetics has found to correlate  most strongly with subjective qualities of pictures (e.g., beauty, creativity, and emotions). 


\item Determining which facial cues people appear\footnote{We say `appear to be using' simply because we measure correlations and, as such, we do not know what people actually use.}  to be using to make their guesses about place ambiance (Section~\ref{sec:person impression}). To this end, we carry out a correlation analysis between the presence of our visual features and ambiance ratings. We find that colors play an important role (e.g., posh and pretentious places are associated with the presence of pink, strange and creative places with that of yellow), and that computational aesthetics features such as uniqueness effectively capture the associations people make with creative places.


\item Showing that our algorithms make  accurate predictions (Section~\ref{sec:algorithm_predictions}). We find that they show a precision error at most of  0.1 on a scale [0,1].

\item Determining which visual cues our algorithms extract from profile pictures  to make their predictions about place ambiance (Section~\ref{sec:algorithm_associations}). We find that they tend to mostly rely on  image quality,  face position, and face pose.


\item Demonstrating that the visual cues people appeared to have relied upon partly differ from those of the algorithm (Section~\ref{sec:people_vs_algorithm}). We find that people rely on emotional associations with colors in expected ways (e.g., green goes with calm places). By contrast,  the algorithms rely on objective features such as those capturing basic compositional rules. We also find that, as opposed to the algorithms, people may be engaging in gender and racial stereotyping, but might be doing so only at times. 
\end{itemize}

Section~\ref{sec:discussion} concludes by discussing this work's limitations, and theoretical and practical implications.

\section{Related Work}
\label{sec:related}


In the context of location-based services, place reviews have been widely explored. Tips and descriptions  have been explored to study the  popularity of Foursquare places~\cite{Vasconcelos:2014:PDF:2660460.2660484,vasconcelos2012tips}; to explain the demand of restaurants in Yelp;  \cite{luca2011reviews}; and to learn their relationship with check-ins and photos at Foursquare venues~\cite{Yu:2014:EOU:2661714.2661724}.



Face images have been studied in different disciplines. Computer vision researchers have analyzed  
faces for several decades. Researchers did so to automatically recognize face ovals~\cite{viola2004robust,turk1991face}, identify face expressions~ \cite{fasel2003automatic}, predict personality traits~\cite{naumann2009personality,biel2013hi}, assess political competence \cite{olivola2010elected},  infer visual persuasion~\cite{Joo_2014_CVPR}, and score portraits for photographic beauty~\cite{redi2015thebeauty}. 

More recently, faces have been also studied in the context of social-networking sites. It has been found that, on Facebook, faces engage users more than other subjects~\cite{bakhshi14faces}, and that  faces partly reflect  personality traits~\cite{haxby2000distributed,mehdizadeh2010self,back2010facebook}.



There has not been any work on using profile pictures of visitors to \emph{automatically} infer the ambiance of commercial places. 



\section{Dataset}
\label{sec:dataset}
To study profile pictures and relate their features to ambiance ratings, we resorted to use the dataset introduced in~\cite{graham11ambience}, which we  briefly summarize next. 

\mbox{ } \\
\textbf{Data for ambiance ratings.} This dataset contains ratings from 49 places on Foursquare (24 bars and 25 cafes) located in Austin, Texas. These establishments were randomly selected and in order to be included into the sample contained at least 25 profile pictures of individuals who frequented the location. 
For each establishment, 72 ambiance dimensions were defined in a way that reflected a place's ambiance (e.g., creepy), clientele (e.g., open-minded), and activities (e.g., pickup). The dataset comes with two sets of ambiance ratings:

\begin{description}
\item \textit{(1) Face-Driven Ambiance Ratings.} Each place's 25 profile pictures were 
arranged on a survey sheet. Ten survey respondents were asked to rate the ambiance of the place (along the 72 dimensions) \emph{only based} on the pictures of its visitors. The respondents were equally distributed across genders (average age: 20.5). The inter-annotator agreement was measured as intra-class agreement and was .32 for ambiance, .69 for clientele, and .33 for activity.

\item \textit{ (2) On-the-Spot Ambiance Ratings.} A separate team of 10 survey respondents were  asked to visit the places and  rate their ambiance on the same 72 dimensions. The participants were, again, equally distributed across genders, and their average age was 22.4 years. The inter-annotator agreement  for on-the-spot ratings was higher: again, measured as intra-class agreement, it was .69 for ambiance, .79 for clientele, and .62 for activity.
\end{description}

The resulting dataset is limited in size because of two main reasons. First,  the number of locations with at least 25 profile pictures is limited. Second, it takes time to have two groups of 5 raters each (research assistants above drinking age)  at every location at two different times of the day/week. Despite that drawback,  for the first time, this dataset thoroughly quantifies the multidimensional ambiance dimensions of real-world locations.

For analytical purposes, it was useful to reduce the 72 ambiance dimensions into a smaller group of dimensions. To this end, we used a semi-automated technique and managed to reduce those ambiance dimensions into 18 orthogonal ones.  We did so upon the dataset of face-driven ambiance ratings (similar results were obtained with the on-the-spot ratings). Each place was expressed with a 72-element vector. We clustered those vectors using $k$-means, set $k$ to one value in  $k=\{5,10,15,20,25,30\}$ at the time,  and obtained different clusters for different values of $k$. To select the best cluster arrangement, we computed the average cluster silhouette~\cite{rousseeuw1987silhouettes}, and found $k=25$ to return the best silhouette. By visual inspection, we determined that $k=25$ indeed returned the best grouping yet this grouping showed some minor inconsistencies (e.g.,  ``trendy'' and ``stylish'' were  assigned to different clusters despite being quite strongly related). To fix that, we run a small user study with 10 participants. We prepared a sheet containing 25 clusters, each of which was described with  ambiance terms (in total, we had 72 terms). A participant had to label every cluster based on its associated terms. If the labeling turned out to be  difficult because of spurious terms, the participant was free to move those terms across clusters. Also, the participant had to underline the most representative term for each cluster (which we call  ``target ambiance'').   The implicit goal of this task was to improve both intra-cluster consistency and inter-cluster diversity.
After analyzing the survey responses, we were left with 18 semantically-consistent ambiance clusters (Table~\ref{table:clusters}) whose target ambiance scores are correlated in expected ways (Figure~\ref{fig:correlation_matrixes}). Face-driven and on-the-spot ratings show very similar correlations.

\begin{table}
\resizebox{\columnwidth}{!}{
\begin{tabular}{ | c | c |p{6cm}|}
\hline
\textbf{definition} & \textbf{target} & \textbf{other ambiances} \\\hline
middle-class&trendy&stylish, modern, white-collar, impress \\\hline
relaxing&relax&cozy, simple, clean, comfortable, pleasant, relaxed, homey \\\hline
posh&formal&luxurious, upscale, sophisticated \\\hline
friendly&cheerful&funny, friendly \\\hline
social&drink /eat&meet new people, watch people, hangout \\\hline
romantic&dating&cheesy, romantic \\\hline
pickup&pickup&meat market \\\hline
creative&artsy&quirk, imaginative, art, eclectic, edgy, unique, hipster, bohemian \\\hline
party&music&energetic, loud, dancing, camp \\\hline
attractive&attractive & \\\hline
open-minded&open&open-minded, adventurous, extraverted \\\hline
blue-collar&blue-collar& \\\hline
traditional&bland&conservative, old-fashion, sterile, stuffy, traditional, politically conservative \\\hline
strange&off path&strange \\\hline
cramp&cramp&dark, dingy, creep \\\hline
calm&agreeable&emotionally stable, concencious \\\hline
reading&read&study, work, web \\\hline
pretentious&douchy&pretentious, self centered \\\hline
\end{tabular} }
\caption{Ambiance clusters and corresponding target ambiance.} \label{table:clusters}
\end{table}

\begin{figure}[t]
\centering
\includegraphics[width=\linewidth ]{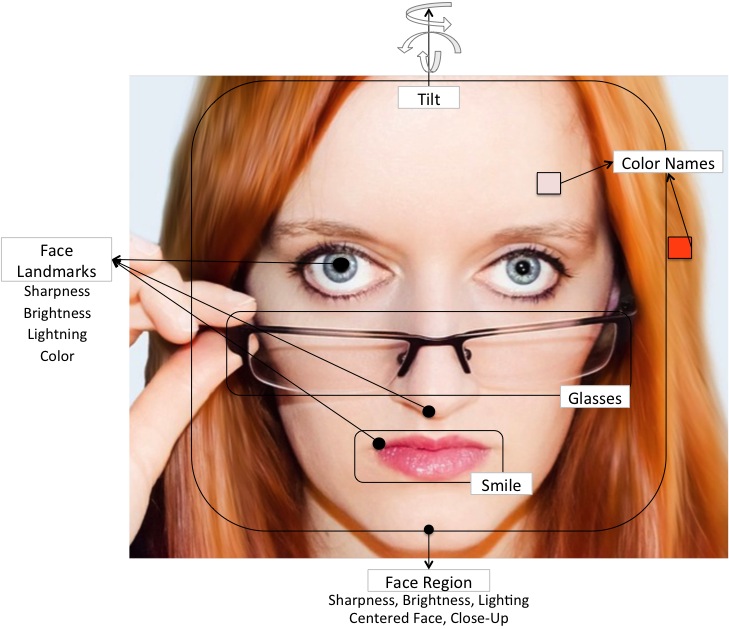}
\caption{Running example of a profile picture.}
\label{fig:portrait}
\end{figure}

\section{Predictors}
\label{sec:predictors}
%
%
%
%
%

Our goal is to predict a place's ambiance from the profile pictures of its visitors. Therefore, next, we need to extract some predictors out of each picture. We cannot use traditional computer vision features as predictors because these are mostly used  to semantically analyze an image, i.e.,  to understand \emph{which}  objects are present in the image~\cite{sivic2003video}. That is not the task at hand simply because, for face-driven ambiance ratings, the main  object of the image is known (it is a  face). By contrast, stylistic characteristics appear more promising. These capture, for example, \emph{how} a face is photographed, its aesthetic value, its dominant colors and textures, its affective content, and  the self-presentation choices it entails. Stylistic features have been used previously to automatically assess images' and videos' aesthetic value\cite{datta,birkhoff1933aesthetic}, expressed emotions \cite{lu2012shape,machajdik2010affective},  creativity \cite{redi6}, and interestingness \cite{gygli2013interestingness}. In a similar way, we collect  here a set of image  features (predictors) that reflect portrait-specific  stylistic aspects of the image, of the face, and of its visual landmarks (eyes, nose, and mouth).  

To infer face demographics, position and landmarks from our profile pictures, we use  Face++, a face analysis software based on deep learning \cite{fan2014learning}. Face++ has been found to be extremely accurate in both face recognition \cite{fan2014learning} and face landmark detection \cite{zhou2013extensive}. The information of whether a face is detected or not is encoded in a binary feature. If a face is not detected (that happened for 47\% of the images), that feature is zero and all face-related features are set to be \emph{missing} values.  Next, we introduce our visual features used as predictors and, to ease illustration, we group them into five main types.
%

%
%
\mbox{} \\
\textbf{1) Aesthetics.} An image style is partly captured by its beauty and  quality. Psychologists have shown that, if the photo of a portrait is of quality, then the portrait is memorable, gives a feeling of familiarity, and better discloses the mood of the subject~\cite{kennedy2009life}. Photo quality has also been related to its level of creativity~\cite{redi6}, and of beauty and interestingness~\cite{gygli2013interestingness}. We implement  computational aesthetics algorithms from \cite{datta,birkhoff1933aesthetic} and score our profile pictures in terms of beauty and quality. More specifically, we compute: 
\begin{description}

\item \emph{Photographic Quality.} The overall visual \textit{photographic quality} reflects the extent to which an image is correct according to standard rules of good photography. To do capture this dimension, we compute the  \textit{camera shake amount} \cite{redi6} (the quantity of blur generated by the accidental camera movements), the \textit{face landmarks sharpness}, and  \textit{face focus} (which has been found to be correlated with beauty \cite{redi2015thebeauty}). To see how those three dimensions translate in practice, consider Figure~\ref{fig:portrait}. This can be considered a good quality picture: there is no camera shake, the face is in focus compared to the background, and the facial landmarks (e.g., eyes, mouth) are extremely sharp.

\item \emph{Brightness, Saturation, Contrast.}  The three aspects respectively correspond to the colors' lightness, colorfulness and discriminability in an entire image. They have all been found to be associated with  picture aesthetic~\cite{datta,redi2015thebeauty} and its affective value~\cite{valdez1994effects}. Darker colors evoke emotions such as anger, hostility and aggression, while increasing brightness evokes feelings of relaxation and is associated with creativity~\cite{redi6}. For each of our pictures, we compute \textit{brightness, lightning, and  saturation} of the eyes, nose, mouth and the entire face oval. To do that, we add an overall \textit{contrast} metric~\cite{redi2015thebeauty}. To stick with our running example, Figure~\ref{fig:portrait} has very bright colors: without being over-saturated (too colorful), the contrast is high enough to make bits of the face quite distinguishable.

\item \emph{Image Order.}  According to Birkhoff, the aesthetic value of a piece of (visual) information can be computed by the ratio between its order (number of regularities) and its complexity (number of regions in which it can be decomposed)~\cite{birkhoff1933aesthetic}. Order and complexity have been found to be associated with beauty,  and to affect how fast humans process visual information~\cite{snodgrass1980standardized}. 
We thus compute the \textit{image order}  and its \textit{complexity} using a few information theory metrics~\cite{redi6}, its \textit{level of detail} (i.e., number of regions resulting after segmentation)~\cite{machajdik2010affective}, and its overall \textit{symmetry}. The picture in  Figure~\ref{fig:portrait} can be considered quite conventional:  lines are symmetric, and regularities are introduced by the uniformity of its background and the smoothness of its textures.

\item \emph{Circles.}  The literature on affective image analysis suggests that  the presence of circular shapes is registered when certain emotions (e.g., anger, sadness) are expressed~\cite{lu2012shape}.
Therefore, we add the \textit{presence of circular shapes} to our list of predictors. We compute them by using Hough's transform~\cite{redi2015thebeauty}. The face in Figure ~\ref{fig:portrait} has perfect round shapes in the eyes area only: 2 for the iris, and 2 for the eye pupils.

\end{description}

\mbox{} \\
\textbf{2) Colors.} They have the power to drive our emotions, and are associated with certain abstract concepts~\cite{mahnke1996color,hemphill1996note,james1953effect}: \emph{red} is related to excitement~\cite{wexner1954degree};  \emph{yellow} is associated to cheerfulness~\cite{wexner1954degree}; \emph{blue} with comfort, wealth, trust, and security~\cite{wexner1954degree}; and \emph{green} is seen as cool, fresh, clear, and pleasing~\cite{mahnke1996color}.
To capture colors from our pictures, we  compute the \textit{color name} features~\cite{machajdik2010affective}  and \textit{facial landmark colors} according to their hue values~\cite{redi2015thebeauty}. In Figure~\ref{fig:portrait}, the dominant colors are white, red, and pinkish.

\mbox{} \\
\textbf{3) Emotions.}  Facial expressions give information not only about the personality of the subjects \cite{biel2012facetube}, but also about the communicative intent of an image \cite{Joo_2014_CVPR}. Faint changes in facial expressions are easily judged by people who often infer  reliable social information from them~\cite{iacoboni01}. That is also because specific areas of the brain are dedicated to the processing of emotional expressions in faces~\cite{Goldman2005193}. We therefore  compute the probability that a face subject assumes one of these emotions: \textit{anger, disgust, happy, neutral, sad}. We do so by resorting to Tanveer \emph{et al.}~\cite{Tanveer:2012:FSS:2384916.2384956}'s work based on eigenfaces~\cite{turk1991face}. We also determine whether a face is \emph{smiling} or not using the  Face++ smile detector.

\mbox{} \\
\textbf{4) Demographics.} The distribution of age and gender among visitors is expected to greatly impact the ambiance of the place. It is well-known that people geographically sort themselves (in terms of where they choose to live, which places they like) depending on their  socio-demographic characteristics and end up clustering with others  who are like-minded~\cite{the-big-sort}. We take \emph{race} (caucasian, black, asian), \emph{age}, and \emph{sex} as our demographic features. 
%
%

\mbox{} \\
\textbf{5) Self-presentation.} The way people present themselves might also be related to what they like~\cite{mehdizadeh2010self}. To partly capture self-presentation characteristics, we  determine  whether \emph{sunglasses} or \emph{reading glasses} are used, 
 whether a picture actually  \emph{shows a face} or not, and, if so, we determine three main facial characteristics:  \emph{face centrality},  whether there is a \emph{tilted face}, and whether it is a \emph{close-up}. Figure~\ref{fig:portrait}, for example, shows a close-up of a non-tilted and centered face. Our last self-presentation feature reflects whether the image composition is unique and memorable (we call it `` \emph{uniqueness}''). It indicates the extent to which the image is novel  compared to the average profile picture~\cite{redi2012interestingness}. \\
%
%
%

To sum up, for each profile picture, we have a total number of   64 features.  To combine the features of a venue's faces together, we  characterize each place with the  \emph{average} and the \emph{standard deviation} of the features across the 25 pictures.
The diversity analysis arising from the standard deviation statistics is needed because ``it seems likely that observers do more than simply averaging the individual impressions of the targets. If targets are too diverse, then the group is seen as diverse \ldots''~\cite{graham11ambience}. For each place, we therefore have a 128-dimensional feature vector, to which we add a value corresponding to the total number of faces present in the group of 25 pictures.  Hence, we represent a place with a final feature vector of  129 elements.

\begin{figure*}[ht!]
\centering
\includegraphics[width=1\linewidth ]{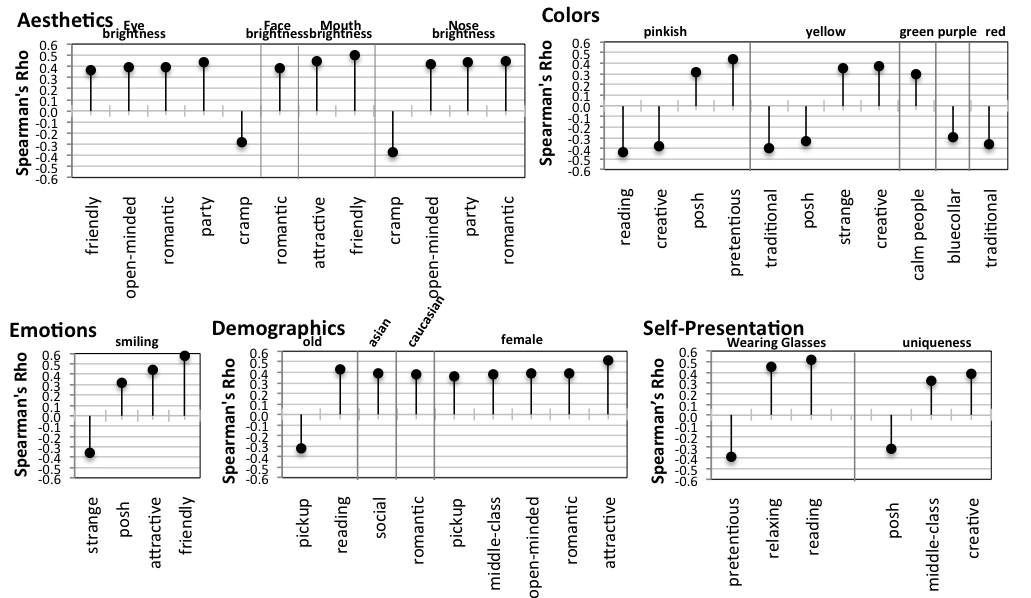}
\caption{Spearman correlations between the patrons' visual features and judgments of where those patrons might like to go. The visual features are grouped into five types: aesthetics, colors, emotions, demographics, and self-presentation features. All the correlations are statistically significant at $p < 0.05$.}
\label{fig:analysis_faces}
\end{figure*}

\section{People Associations}
\label{sec:person impression}

To determine which visual cues the respondents in our dataset may have relied upon to make their guesses, we study the extent to which a person's visual features impacted respondents' guesses. We resort to the face-driven ambiance ratings introduced in Section~\ref{sec:dataset}. For each place, we compute the pairwise correlations between each of the 129 visual features and each the 18  ambiance ratings. Of course, face-specific features are defined only for images that contain faces. To compute those 2,322 correlations, we use the  Spearman Correlation as most of the visual features reflect the presence or absence of visual elements (e.g., glasses) and, as such, they are best interpreted in a comparative fashion rather than using raw numbers. To ease illustration, next, we will group the correlation results (Figure~\ref{fig:analysis_faces}) by type of features. \\

\begin{description}
\item  \textbf{Aesthetics Features.} The most relevant aesthetic feature is brightness. Respondents associate eye brightness, mouth brightness, and nose  brightness with people who like  friendly, open-minded, romantic, and party places (with all correlations above $r=0.4$). By contrast, dark pictures - in terms of face brightness ($r=-0.28$) and nose brightness ($r=-0.37$) - are associated with those who like cramp places.

\item \textbf{Colors.} The presence of pink in profile pictures is associated with those who like posh and calm places, while its absence is associated with those who like reading and creative places. The presence of yellow is associated  with strange  and creative people, and its absence with those who like traditional and posh places. 

\item \textbf{Emotions.}  The most important emotion feature is smiling. Profile pictures with smiling faces are associated with those who like posh, attractive, and friendly places (smiling faces are associated with friendly places with a correlation as high as $r=0.57$), while strange people are thought not to smile. 

\item \textbf{Demographics.} Old people are  associated with reading places but, of course, not with pickup places. Race is also associated with ambiance: Asian are associated with social places, while Caucasian with romantic places. The presence of female among a place's visitors results into considering the place to be good for pickup, to be middle-class, open-minded, romantic, and catered to attractive people.

\item \textbf{Self-Presentation.} Those who wear glasses are associated with relaxing and reading places, while those who do not with pretentious places. Those who use profile pictures that deviate from the conventional one (i.e., they tend to be \emph{unique}) are associated with middle-class and creative places, while those who are conventional are associated with posh places. 

\end{description}



\section{Algorithmic Predictions}
\label{sec:algorithm_predictions}

We have just seen that survey respondents made systematic associations between place ambiance and visual features. Now one might wonder whether an algorithm could make associations as well to automatically predict place ambiance ratings. 

To address that question, we use the on-the-spot ratings (Section~\ref{sec:dataset}). Hence, for each place, we have 72 ambiance ratings (which might well differ from the face-induced analyzed in the previous section). Again, those ambiance ratings are summarized into 18.

Having this data at hand, we could train a regression framework on part of the 49 places (each of which is represented by the usual 129 features), and we could then test the extent to which the framework is able to predict the 18 ambiance dimensions on the remaining places.  The problem is that we have too few observations. To avoid overfitting,  the standard rule for regression is to have at least 10 observations per variable \cite{peduzzi1996simulation}. Unfortunately, in our case, we have 49 places  (observations) and 129 features (variables).

To fix this problem,  we train the regression framework not on all the 129 features but on the 5 most correlated ones. Since our dataset consists of only 49 observations, we need to carefully maximize the number of training samples. Even a 90 \%-10\% train-test  partition might  be restrictive as it might  remove important outliers from the training data. To tackle this issue, we resort to the widely-used leave-one-out validation~ \cite{salakhutdinov2013learning}. At each iteration, we leave one sample out as test, and train the framework on the remaining samples; the job of the framework is to predict the test sample. The difference between the predicted value and the actual one (that was left out) is the error, which we summarize as the percentage Mean Squared Error (MSE). 

\begin{figure}[t]
\centering
\includegraphics[width=0.9\linewidth ]{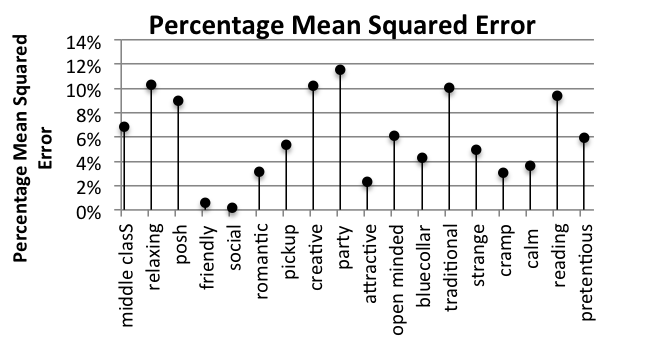}
\caption{Accuracy Errors of the Algorithmic Predictions. 
}
\label{fig:results_places}
\end{figure}

Figure~\ref{fig:results_places} shows  that our framework accurately predicts all the ambiance dimensions (the error is always below 10\%) despite having only five visual features as input. The two ambiance dimensions of friendly and social are particularly easy to predict (the error is close to 0). That is because the pictures in those places tend to be distinctive: they are similar to each other but differ from the pictures of other ambiance dimensions. Party places tend to be associated with relatively more diverse pictures, yet the error is quite reasonable (12\%).

\section{Algorithmic Associations}
\label{sec:algorithm_associations}

\begin{figure*}[t]
\centering
\includegraphics[width=1\linewidth ]{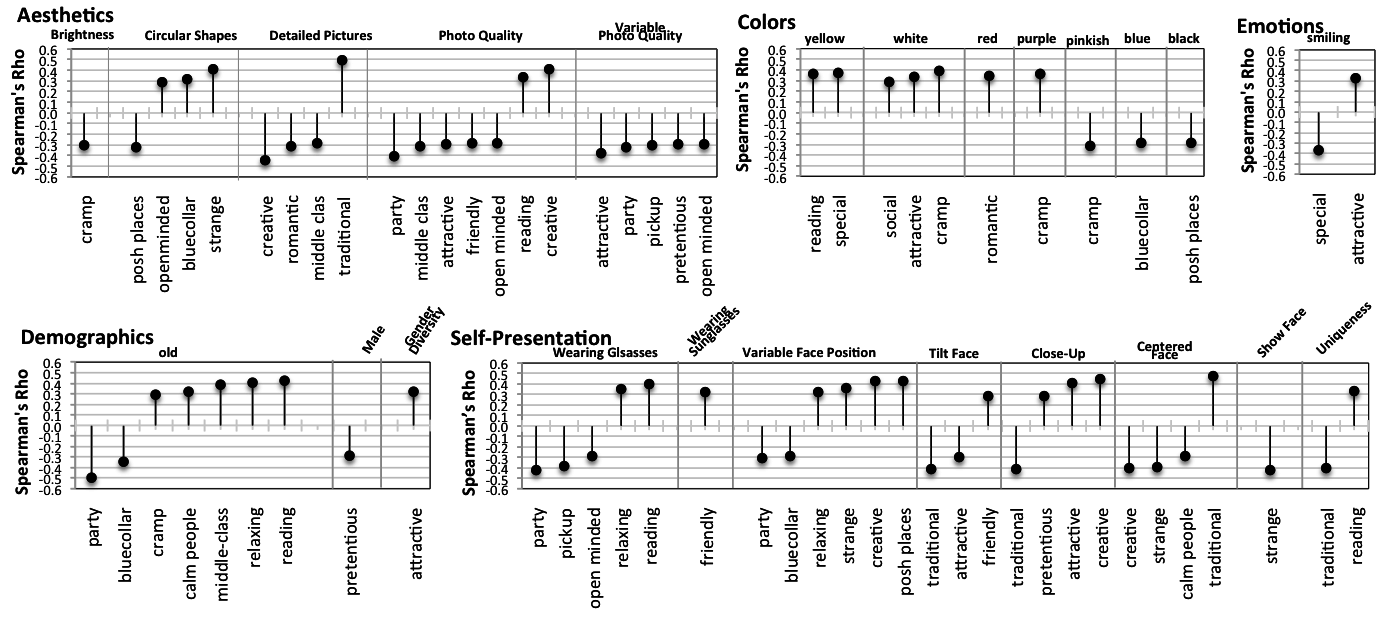}
\caption{Spearman correlations between visual features of a place's visitors and the place's actual ambiance. The visual features are grouped into five types: aesthetics, colors, emotions, demographics, and self-presentation features. All the correlations are statistically significant at $p < 0.05$. 
}
\label{fig:analysis_places}
\end{figure*}

So the framework makes accurate predictions of place ambiance, suggesting that visual features are not only likely impact people's perceptions (as one would expect from the literature) but are indeed related to the places one likes: one's face  partly reveals one's likes. To determine which visual cues our algorithm has used to make its predictions, for each place, we compute the  correlation between each of the 129 visual features and each the 18 actual ambiance ratings. To compute those 2,322 correlations, we use, again, the  Spearman Correlation. By grouping the correlation results by feature type (Figure~\ref{fig:analysis_places}), we see that: 

\begin{description}

\item \textbf{Aesthetics Features.} Dark pictures (those lacking brightness) are indeed used by people who go to cramped places. That is in line with the associations made by the respondents. Our framework finds circular shapes in the profile pictures of people who go to open-minded, blue-collar, and strange places, while they are absent in the profiles of those who go to posh places. Unlike those going to creative, romantic and middle-class places, those going to traditional places tend to use pictures with more complex backgrounds. People going to relaxing and creative places use quality pictures in their profiles. By contrast, people going to party, middle-class, attractive, friendly, and open-minded places are less prone to quality images:  in any of those places, not only a few  pictures are of low quality but, since the group of pictures as a whole shows low variability (see panel `variable photo quality' in Figure~\ref{fig:analysis_places}), all of the place's pictures are systematically of low quality. 

\item \textbf{Colors.} The profile pictures making use of yellow are of  those going to relaxing and strange places; white of those going to social, attractive, and cramped places; red of those going to romantic places; purple of those going to cramped places. Instead, pinkish pictures are avoided by those going to cramped places, blue ones by those going to blue-collar places, and black  ones by those going to posh places.

\item \textbf{Emotions.}  The most important emotion feature is, again,  smiling. In line with the respondents' assessments,  our framework learned that those going to strange places do not smile, while those going to places catered to attractive people do so. 

\item \textbf{Demographics.} Old people do not go to party and blue-collar places, but they are found in cramped, calm, middle-class, and relaxing places. Men seem to avoid pretentious places, while a balanced woman-man ratio  tends to be enjoyed by  places where attractive people are thought to go. 

\item \textbf{Self-Presentation.} Interestingly, the self-presentation features that matter the most boil down to only two: the use of glasses and face position. Those who wear glasses go to relaxing places, while those who do not wear them go to party, pickup, and open-minded places. People wearing sunglasses go to friendly places. As for face position, those going to  blue-collar and party places tend to have their faces in a similar position, while a variety of positions is experimented by those going to relaxing, strange, creative and posh places. Those going to friendly places tilt their heads. By contrast,  those going to traditional places  do not do so: their faces are centered, and they avoid close-ups. Instead, those going to creative and pretentious places indulge in close-ups.  In addition to not smiling, strange people seems to have a tendency to not always show their faces. Finally, the uniqueness feature also matters: those who use profile pictures that deviate from the conventional one go to reading places, while those who have conventional pictures go to traditional places.

\end{description}

\begin{table*}
 \small
\resizebox{\linewidth}{!}{
\begin{tabular}[t]{| l |  c | c |c|p{8cm}|p{7cm}|p{3.5cm}|}
\hline
\multirow{2}{*}{\textbf{Ambiance}} & \multirow{2}{*}{\textbf{People}}& \multirow{2}{*}{\textbf{Algorithm}} &  &  \multicolumn{3}{c |} {\textbf{Top-5 Features}} \\\cline{4-7}  
&&& &\multicolumn{1}{c}{\textbf{People} }& \multicolumn{1}{c}{\textbf{Algorithm} }& Both \\\Xhline{3\arrayrulewidth}
\multirow{2}{*}{\textbf{middle class}} & \multirow{2}{*}{{0.26}}& \multirow{2}{*}{\textbf{0.45**}} & 
	$\uparrow$ &uniqueness, female&old&  \\\cline{4-7}  
&&& $\downarrow$&male& & detail, photo quality   \\\Xhline{2\arrayrulewidth}
\multirow{2}{*}{\textbf{relaxing}} & \multirow{2}{*}{\textbf{0.45**}}& \multirow{2}{*}{\textbf{0.45**}} & 
	$\uparrow$ &&old& glasses, glasses (variability)  \\\cline{4-7}  
&&& $\downarrow$&show face& variable position&  \\\Xhline{2\arrayrulewidth}
\multirow{2}{*}{\textbf{posh}} & \multirow{2}{*}{{0.15}}& \multirow{2}{*}{\textbf{0.30*}} & 
	$\uparrow$ &smile, pinkish&variable position&  \\\cline{4-7}  
&&& $\downarrow$&yellow, quality, unique&black (variability), centered face, number of circles &  \\\Xhline{2\arrayrulewidth}
\multirow{2}{*}{\textbf{friendly}} & \multirow{2}{*}{{-0.02}}& \multirow{2}{*}{\textbf{0.42**}} & 
	$\uparrow$ &smile, brightness (mouth,eyes)&sunglasses, sunglasses (variability)&  \\\cline{4-7}  
&&& $\downarrow$&photo quality (variability)&brightness (mouth, nose) (variability)&  \\\Xhline{2\arrayrulewidth}
\multirow{2}{*}{\textbf{social}} & \multirow{2}{*}{{0.15}}& \multirow{2}{*}{\textbf{0.30*}} & 
	$\uparrow$ &asian, asian (variability), caucasian (variability)&white, white(std), sad (std), brightness (face) (variability)& \\\cline{4-7}  
&&& $\downarrow$&&& \\\Xhline{2\arrayrulewidth}
\multirow{2}{*}{\textbf{romantic}} & \multirow{2}{*}{{0.15}}& \multirow{2}{*}{\textbf{0.29*}} & 
	$\uparrow$ &brightness (nose, mouth, eyes), female, caucasian&color (nose, eye), color (nose, eye) (variability), red&  \\\cline{4-7}  
&&& $\downarrow$&&&  \\\Xhline{2\arrayrulewidth}
\multirow{2}{*}{\textbf{pickup}} & \multirow{2}{*}{\textbf{0.43**}}& \multirow{2}{*}{{0.34*}} & 
	$\uparrow$ &female, photo quality&&  \\\cline{4-7}  
&&& $\downarrow$&male, old&glasses, glasses (variability), photo quality&  \\\Xhline{2\arrayrulewidth}
\multirow{2}{*}{\textbf{creative}} & \multirow{2}{*}{\textbf{0.60***}}& \multirow{2}{*}{{0.45**}} & 
	$\uparrow$ &uniqueness (variability), uniqueness, yellow, white (variability)&close-up&  \\\cline{4-7}  
&&& $\downarrow$&pinkish&centered face, detail&  \\\Xhline{2\arrayrulewidth}
\multirow{2}{*}{\textbf{party}} & \multirow{2}{*}{\textbf{0.58***}}& \multirow{2}{*}{{0.44**}} & 
	$\uparrow$ &female&&  \\\cline{4-7}  
&&& $\downarrow$& &glasses, old& photo quality (variability), glasses (variability) \\\Xhline{2\arrayrulewidth}
\multirow{2}{*}{\textbf{attractive}} & \multirow{2}{*}{{0.39***}}& \multirow{2}{*}{\textbf{0.57***}} & 
	$\uparrow$ &female, brightness (mouth, face)&closeup, white, white (variability), gender diversity&smile  \\\cline{4-7}  
&&& $\downarrow$&male&photo quality&  \\\Xhline{2\arrayrulewidth}
\multirow{2}{*}{\textbf{open-minded}} & \multirow{2}{*}{\textbf{0.50***}}& \multirow{2}{*}{{0.18}} & 
	$\uparrow$ &brightness (mouth, face, eyes)&presence of circles&  \\\cline{4-7}  
&&& $\downarrow$&male&glasses, photo quality&  \\\Xhline{2\arrayrulewidth}
\multirow{2}{*}{\textbf{bluecollar}} & \multirow{2}{*}{{0.10}}& \multirow{2}{*}{\textbf{0.41**}} & 
	$\uparrow$ &&presence of circles&  \\\cline{4-7}  
&&& $\downarrow$&sad, sad (variability), photo quality, purple&old, blue, face position (variability)&  \\\Xhline{2\arrayrulewidth}
\multirow{2}{*}{\textbf{traditional}} & \multirow{2}{*}{\textbf{0.59***}}& \multirow{2}{*}{{0.46***}} & 
	$\uparrow$ &&centered face&detail  \\\cline{4-7}  
&&& $\downarrow$&yellow, red, yellow (variability), uniqueness (variability)&&  \\\Xhline{2\arrayrulewidth}
\multirow{2}{*}{\textbf{strange}} & \multirow{2}{*}{{0.41**}}& \multirow{2}{*}{\textbf{0.56***}} & 
	$\uparrow$ &yellow (variability), centered face, close-up (variability)&black (variability), presence of circles& yellow  \\\cline{4-7}  
&&& $\downarrow$&&shows face, centered face&  smile\\\Xhline{2\arrayrulewidth}
\multirow{2}{*}{\textbf{cramp}} & \multirow{2}{*}{{0.31*}}& \multirow{2}{*}{\textbf{0.46***}} & 
	$\uparrow$ &disgusted (variability)&white, white (variability), purple, purple (variability)&  \\\cline{4-7}  
&&& $\downarrow$&&& brightness (nose, eyes) \\\Xhline{2\arrayrulewidth}
\multirow{2}{*}{\textbf{calm}} & \multirow{2}{*}{{0.41**}}& \multirow{2}{*}{\textbf{0.24}} & 
	$\uparrow$ &centered face, green, green (variability)&old&  \\\cline{4-7}  
&&& $\downarrow$&&centered face (variability)&  \\\Xhline{2\arrayrulewidth}
\multirow{2}{*}{\textbf{reading}} & \multirow{2}{*}{\textbf{0.78***}}& \multirow{2}{*}{{0.59***}} & 
	$\uparrow$ & glasses (variability), photo quality, photo quality (variability)&yellow&  glasses, old\\\cline{4-7}  
&&& $\downarrow$&pinkish&&  \\\Xhline{2\arrayrulewidth}
\multirow{2}{*}{\textbf{pretentious}} & \multirow{2}{*}{{0.33*}}& \multirow{2}{*}{{0.28}} & 
	$\uparrow$ &pinkish&presence of circles&  \\\cline{4-7}  
&&& $\downarrow$&glasses (variability), glasses&photo quality(variability), male&  \\\Xhline{2\arrayrulewidth}
	\end{tabular} \label{tab:comparison}}
 \caption{Predictive accuracy of the ambiance dimensions for the group of people vs. the algorithm. (Note: *** = p $<$ .01; ** = p $<$ .01; *= p$< $.05)}
 \end{table*}

\section{People \emph{vs.} Algorithmic Associations}
\label{sec:people_vs_algorithm}

We have seen that both the algorithmic framework and the group of people are able to estimate the  ambiance of a place given the profile pictures of its visitors. One might now wonder who is more accurate: the algorithms or the group of people? The answer is `depends'. To see why, consider Table~\ref{tab:comparison}. A row in it refers to a specific ambiance dimension and reports the predictive accuracy (Spearman correlation) for our group of respondents ($2^{nd}$ column) and that for our algorithmic framework ($3^{rd}$ column). The highest accuracy between the two is marked in bold. The remaining three columns report the top5 features (if available) that turn out to be most important for people, for the algorithm, and for both. The features are placed in two rows depending on whether they are correlated positively ($\uparrow$ row) or negatively ($\downarrow$ row). 

The algorithm performs better in half of the cases. It generally does so for ambiance dimensions that are well-defined (e.g., posh, friendly, social, romantic). For more multi-dimensional and complex dimensions (e.g., creative, open-minded), the group of people outperforms, but not to a great extent (the predictive accuracies are quite comparable). 

\mbox{ } \\
\textbf{Algorithm wins.} As opposed to the algorithm, people find it difficult to correctly guess these five ambiance dimensions from profile pictures (the correlations are below $0.20$): 
\begin{description}
\item \emph{Posh places.}  People appear to rely on the presence of pinkish and yellow, and that of smile; the algorithm, instead, relies on the presence of black and of circular shapes, and on the face position;

\item  \emph{Friendly places.} People are likely to rely on smiles, while the algorithm relies on sunglasses (and  achieves higher accuracy in so doing).

\item  \emph{Social places.} People seems to engage in race profiling, while the algorithm goes more for picture brightness and presence of the white color. 

\item  \emph{Romantic places.} People look at the presence of women and of bright pictures, while the algorithm look for red elements and for warmer colors in the face landmark. 

\item  \emph{Blue-collar places.} People unsuccessfully go for the absence of purple, while the algorithm successfully goes for the absence of blue. 
\end{description}

\mbox{ } \\
\textbf{People win.}  As opposed to people, the algorithm finds it difficult to correctly guess one ambiance dimension from profile pictures: that of  \emph{open-minded places.}  The algorithm relies on the presence of circular shapes and on the absence of reading glasses. Instead, people correctly infer that very bright pictures are typical for open-minded places.


\mbox{ } \\
\textbf{They both agree.} More interestingly, from the last column of Table~\ref{tab:comparison}, we see that the algorithm partly relies on people's  stereotyping: both agree that the profile pictures for middle-class places suffer from the lack of details (as opposed to those for traditional places); those for relaxing and reading places portray elderly people with reading glasses; those for places catered to attractive people have smiling faces; those for cramped places show darker lightning; and those for strange places show high variability in the use of yellow.

\section{Discussion}
\label{sec:discussion}
We have shown that a state-of-the-art algorithmic framework is able to predict the ambiance of a place based only on 25 profile pictures of the place's visitors. By then looking at the  ambiance-visuals associations that the framework has made, we have tested to which extent those associations reflect people's stereotyping. 



\mbox{ } \\
\textbf{Limitations and Generalizability.} Some of our results reflect what one expects to find (e.g., the color green is associated with calm places), and  that speaks to the validity of those results (concurrent validity). However, one of the limitations of this work is still that we cannot establish how our insights generalize beyond restaurants and cafes in Austin. It is well-known that ambiance and people perceptions change across countries. To conduct such a study, new sets of data need to be collected, perhaps by combining the use of Mechanical Turk (to get the face-driven ratings) and that of TaskRabbit (to get the on-the-spot ratings). Also, researchers in the area of social networks might complement our study and analyze the relationship between activity features (e.g., reviews, ratings) and ambiance.

\mbox{ } \\
\textbf{Theoretical Implications.} We have also contributed to the literature of computational aesthetics. Some of our visual features have been indeed used in previous studies that automatically  scored  pictures for beauty and emotions~\cite{datta,machajdik2010affective}. We have now learned that the very same features can help infer place ambiance. More broadly, if one thinks about those places within the  context of neighborhoods, then our framework could be used to infer ambiance of not only  individual places but also of entire cities.  Financial and social capitals of  neighborhoods have been widely explored in economics, and aesthetic capital (i.e., the extent to which neighborhoods are perceived to be beautiful and make people happy) has also received attention~\cite{quercia2014aesthetic}. Adding the concept of ambiance capital to current urban studies might open up an entirely new stream of research in which human perceptions take center stage.

\mbox{ } \\
\textbf{Practical Implications.} Our results might enable a number of applications. The easiest one 
is to recommend places to users  based on ambiance. People increasingly manage their private parties by creating  Facebook invitation pages. Such a page contains the profile pictures of those who are invited, and of those who are definitely going. To that group of people, it might now be possible to recommend the party venue that best matches those people's ambiance preferences. It is also possible to do the opposite: to  recommend the best faces for a business (e.g., the faces to show on the web page of a coffee shop). To see why  the business world might be interested in that, consider the work done by the sociologist Yasemin Besen-Cassino. She found that ``chains target affluent young people, marketing their jobs as cool, fashionable, and desirable \ldots Soon, their workers match their desired customers.''~\cite{cassino13cool} Indeed, according to the 2000 Department of Labor's Report on Youth Labor Force, youth from higher socio-economic status are more likely to work than their less affluent counterparts. As Besen-Cassino puts it: ``young [affluent] people see low-paid chain sotres as places to socialze with friends away from watchful parental eyes. They can try on adult roles and be associated with their favorite brands.''~\cite{cassino13cool}


\mbox{ } \\
\textbf{Legible  Algorithms.} We wish to stress that this descriptive study was possible only because we have used visual features that are explainable. Increasingly, the area of computer vision is moving away from descriptive features,  and researchers are putting their effort into high-accuracy  deep-learning algorithms~\cite{bengio2009learning}. Yet, here we have shown that, despite the small quantity of input data, considerable accuracy has been enjoyed by state-of-the-art vision approaches that do not rely on black-box features. Those approaches might be beneficial to 
a broad audience of researchers (e.g., computational social scientists) who need interpretable insights.

\section{Conclusion}



Faces are telling, and researchers have known that for decades. We have re-framed the importance of faces within the context of social-networking sites, and we have shown surprisingly strong associations between faces and ambiance. Beyond the practical implications of our insights, the theoretical ones are of interest: we have seen that what appears to be people's stereotyping matches objective statistical insights at times (e.g., reading glasses go with reading and relaxing places), while it is unfounded at other times (when, e.g., dealing with race and gender). 

We are currently exploring a variety of ways to collect more sets of data at scale. The goal behind this effort is to rephrase this ambiance work within the wider urban context.

\balance

\bibliography{sample}

\bibliographystyle{aaai}
\end{document}